\documentclass[aps, onecolumn, superscriptaddress, 12pt]{revtex4}
\usepackage{amssymb}
\usepackage{amsmath}
\usepackage{bm}
\usepackage[dvips]{graphicx}
\usepackage{dcolumn}
\usepackage{longtable}
\usepackage{color}
\usepackage[dvipdfm]{hyperref}
\def\vect#1{\mbox{\boldmath $#1$}}
\usepackage{ulem}

\begin{document}

\title{Controlling the helicity of magnetic skyrmions in a $\beta$-Mn-type high-temperature chiral magnet}
\author{K.~Karube}
\affiliation{RIKEN Center for Emergent Matter Science (CEMS), Wako 351-0198, Japan.}
\author{K.~Shibata}
\affiliation{RIKEN Center for Emergent Matter Science (CEMS), Wako 351-0198, Japan.}
\author{J.~S.~White}
\affiliation{Laboratory for Neutron Scattering and Imaging (LNS), Paul Scherrer Institute (PSI),
CH-5232 Villigen, Switzerland.}
\author{T.~Koretsune}
\affiliation{Department of Physics, Tohoku University, Sendai 980-8578, Japan.}
\author{X.~Z.~Yu}
\affiliation{RIKEN Center for Emergent Matter Science (CEMS), Wako 351-0198, Japan.}
\author{Y.~Tokunaga}
\affiliation{Department of Advanced Materials Science, University of Tokyo, Kashiwa 277-8561, Japan.}
\author{H.~M.~R\o nnow}
\affiliation{Laboratory for Quantum Magnetism (LQM), Institute of Physics, \'Ecole Polytechnique F\'ed\'erale de Lausanne (EPFL), CH-1015 Lausanne,
Switzerland.}
\author{R.~Arita}
\affiliation{RIKEN Center for Emergent Matter Science (CEMS), Wako 351-0198, Japan.}
\affiliation{Department of Applied Physics, University of Tokyo, Bunkyo-ku 113-8656, Japan.}
\author{T.~Arima}
\affiliation{RIKEN Center for Emergent Matter Science (CEMS), Wako 351-0198, Japan.}
\affiliation{Department of Advanced Materials Science, University of Tokyo, Kashiwa 277-8561, Japan.}
\author{Y.~Tokura}
\affiliation{RIKEN Center for Emergent Matter Science (CEMS), Wako 351-0198, Japan.}
\affiliation{Department of Applied Physics, University of Tokyo, Bunkyo-ku 113-8656, Japan.}
\author{Y.~Taguchi}
\affiliation{RIKEN Center for Emergent Matter Science (CEMS), Wako 351-0198, Japan.}

\begin{abstract}
\:
Magnetic helices and skyrmions in noncentrosymmetric magnets are representative examples of chiral spin textures in solids. 
Their spin swirling direction, often termed as the magnetic helicity and defined as either left-handed or right-handed, is uniquely determined by the Dzyaloshinskii-Moriya interaction (DMI) in fixed chirality host crystals.
Thus far, there have been relatively few investigations of the DMI in metallic magnets as compared with insulating counterparts.
Here, we focus on the metallic magnets Co$_{8-x}$Fe$_x$Zn$_8$Mn$_4$ (0 $\leq$ $x$ $\leq$ 4.5) with a $\beta$-Mn-type chiral structure and find that as $x$ varies under a fixed crystal chirality, a reversal of magnetic helicity occurs at $x_\mathrm{c}$ $\sim$ 2.7.
This experimental result is supported by a theory based on first-principles electronic structure calculations, demonstrating the DMI to depend critically on the electron band filling. Thus by composition tuning our work shows the sign change of the DMI with respect to a fixed crystal chirality to be a universal feature of metallic chiral magnets.
\end{abstract}

\maketitle

\section{Introduction}

Chirality of matter, described as left-handed or right-handed, is an important concept that permeates diverse areas of science, ranging from physics, chemistry to biology. 
In condensed-matter physics, chirality can be displayed by both crystal structures and symmetries, and magnetic spin textures.
One example is magnetic helices in structurally chiral materials where magnetic moments gradually rotate in a clockwise (CW) or counterclockwise (CCW) manner along a helical axis. 
A further related example is magnetic skyrmions, vortex-like topological spin textures\cite{Bogdanov,Nagaosa}. 
Thus far, these magnetic textures have been extensively studied for several classes of chiral magnets\cite{Muhlbauer,Yu_FeCoSi,Milde,Kanazawa_MnGe,Tanigaki,Yu_FeGe,Pappas,Qian,Pollath}. 
Among them, skyrmions exhibit various intriguing properties, and are anticipated to be applied to magnetic memory devices since they can be treated as particles and can be driven by a low current density\cite{Jonietz,Schulz,Iwasaki,Sampaio}.

A microscopic mechanism producing helices and skyrmions in these chiral magnets is a competition between ferromagnetic exchange interaction $J$ and Dzyaloshinskii-Moriya interaction (DMI) $D$, the latter of which arises from the relativistic spin-orbit interaction. 
The effect of the DMI is to gradually twist otherwise ferromagnetically-coupled magnetic moments to form a helix with a long period (10-100 nm) as described by a magnetic wavevector $\vect{q}$.
Here, $|\vect{q}| (= q)$ is proportional to $D/J$, and the helical periodicity $\lambda$ is given by $\lambda = 2\pi/q$.
Depending on the swirling direction of the magnetic moments, magnetic helicity is defined as either left-handed or right-handed, as shown in Figs. 1(c) and (d).
Near the helimagnetic transition temperature $T_\mathrm{c}$, magnetic fields induce a triangular-lattice skyrmion crystal (SkX), which is often described as a triple-$\vect{q}$ structure with the $\vect{q}$ vectors displaying mutual 120$^\circ$ angles.
Figures 1(e) and (f) show that the swirling direction of the in-plane magnetic moments in the skyrmions is either CW or CCW, with the helicity of the magnetic helices and skyrmions being one-to-one correspondent with each other as illustrated in Figs. 1(c-f). 
It is established that the magnetic helicity is uniquely determined by the sign of $D$\cite{Grigoriev_MnFeSi,Grigoriev_FeCoSi,Morikawa_MnSi,Grigoriev_MnFeGe,Shibata,Grigoriev_FeCoGe,Dyadkin}, while the size of a skyrmion is governed by the magnitude of $D$. Therefore, for a versatile use of skyrmions in applications it is important to have a control over $D$.

Previous experimental studies revealed that the relative sign of $D$ can be different within B20-type compounds with the space group of $P$2$_1$3, as for example, between Mn$_{1-x}$Fe$_x$Si (0 $\leq$ $x$ $\leq$ 0.11) and Fe$_{1-x}$Co$_{x}$Si (0 $\leq$ $x$ $\leq$ 0.5)\cite{Grigoriev_MnFeSi,Grigoriev_FeCoSi,Morikawa_MnSi}.
Remarkably, a continuous variation in $D$ as a function of chemical composition, and even its sign reversal were found in Mn$_{1-x}$Fe$_x$Ge\cite{Grigoriev_MnFeGe,Shibata}; at a critical composition $x_\mathrm{c}$ $\sim$ 0.8, $D$ is almost zero and the magnetic ground state is nearly ferromagnetic ($q$ $\sim$ 0) rather than helimagnetic. 
In Fe$_{1-x}$Co$_x$Ge\cite{Grigoriev_FeCoGe}, the sign reversal of $D$ is reported to occur around $x_\mathrm{c}$ $\sim$ 0.6.
Theoretically, it was a challenge to describe the DMI on the long-period, smoothly varying magnetization textures, particularly in the B20-type metallic magnets. 
Recently, however, there has been a significant progress in the theoretical approaches, such as those based on the total energy of spin spirals\cite{Heide_PRB,Heide_PhysicaB,Gayles}, tight-binding representation\cite{Katsnelson}, Berry phase\cite{Gayles,Freimuth_PRB,Freimuth_JPCM}, multi-scale approach with energy mapping\cite{Janson}, spin susceptibility\cite{Koretsune}, or spin current\cite{Kikuchi} formulations.
Promisingly, the aforementioned variation in DMI accompanied by sign reversal can be understood in terms of a change in band filling of 3$d$ orbitals\cite{Gayles} within a complex electronic structure with multiple band-anticrossing points\cite{Koretsune,Kikuchi}. 
Nevertheless, it remains elusive whether such a sign change in DMI may take place universally in other metallic magnets with different chiral crystal structures and moreover with higher $T_\mathrm{c}$s that lie closer to room temperature.
In this paper, we report experimental and theoretical investigations of $\beta$-Mn-type Co-Fe-Zn-Mn alloys, and reveal how the DMI varies as a function of band-filling in a class of metallic chiral magnet very much distinct to the B20-type compounds. 

Co-Zn-Mn alloys crystallize in a $\beta$-Mn-type chiral cubic structure with a space group $P$4$_1$32 or $P$4$_3$32, where 20 atoms per unit cell are distributed over two Wyckoff sites (8$c$ and 12$d$)\cite{Xie,Hori}, as illustrated in Figs. 1(a) and (b).
Recently, Co-Zn-Mn alloys have been found to host DMI-induced helices and skyrmions at and above room temperature\cite{Tokunaga}. 
Mn-free Co$_{10}$Zn$_{10}$ shows a helimagnetic state with $\lambda$ $\sim$ 185 nm below $T_\mathrm{c}$ $\sim$ 460 K, and both $\lambda$ and $T_\mathrm{c}$ decrease as partial substitution of Mn proceeds. 
In Co$_8$Zn$_8$Mn$_4$ ($\lambda$ $\sim$ 120 nm, $T_\mathrm{c}$ $\sim$ 300 K), a SkX state forms at room temperature under magnetic fields as an equilibrium state, and over a much wider temperature and field region as a metastable state\cite{Karube_884}. 
Further increasing the Mn concentration suppresses the helimagnetic state, and instead promotes a spin glass state, probably due to frustrated antiferromagnetic correlations of Mn moments\cite{Karube_776}.
While it was reported that $D$ in Co$_8$Zn$_8$Mn$_4$ and Co$_9$Zn$_9$Mn$_2$ are several times smaller than that in FeGe\cite{Takagi}, no systematic investigation of the DMI as a function of band filling has been performed yet. 

Here, we target the Fe-doped system Co$_{8-x}$Fe$_x$Zn$_8$Mn$_4$, where a single Fe substitution for a Co corresponds to removing one electron per unit cell from the system.
Bulk samples were successfully synthesized while keeping the $\beta$-Mn-type structure and spatial homogeneity of the element distribution up to $x$ = 4.5. 
To clarify the magnetic and crystal structures, we have performed measurements of magnetization, ac magnetic susceptibility and small-angle neutron scattering (SANS) on the bulk samples, as well as convergent-beam electron diffraction (CBED) and Lorentz transmission electron microscopy (LTEM) on the thin-plate specimens.
CBED measurements can directly determine the crystal chirality, namely space group $P$4$_3$32 (defined as left-handed crystal with $\Gamma_\mathrm{c} = -1$) or $P$4$_1$32 (defined as right-handed crystal with $\Gamma_\mathrm{c} = +1$).
On the other hand, LTEM measurements can unambiguously identify the helicity of skyrmions (CCW or CW), thereby allowing the determination of helicity of the corresponding helices, namely left-handed helix ($\gamma_\mathrm{m}$ = $-$1) or right-handed helix ($\gamma_\mathrm{m}$ = $+$1) as shown in Figs. 1(c-f). 
Here, when $\vect{q}$ is parallel (anti-parallel) to $\vect{m}_1 \times \vect{m}_2$, $\vect{m}_1$ and $\vect{m}_2$ being magnetic moments in order along $\vect{q}$, the magnetic helicity is defined as right-handed (left-handed). 
As discussed in previous studies\cite{Grigoriev_MnFeSi,Grigoriev_FeCoSi,Morikawa_MnSi,Grigoriev_MnFeGe,Shibata,Grigoriev_FeCoGe,Dyadkin}, the sign of the DMI is represented by the correlation $\Gamma_\mathrm{c}\gamma_\mathrm{m}$ due to both crystal chirality and magnetic helicity.
From these multiple experiments, we found that the DMI systematically varies with the change in band filling and its sign change occurs around $x_\mathrm{c}$ $\sim$ 2.7 in Co$_{8-x}$Fe$_x$Zn$_8$Mn$_4$.
Finally, the experimental results were compared to the theoretical results based on \textit{ab initio} electronic structure calculation.

\section{Methods}

\subsection{Sample preparation} 

Bulk polycrystalline ingots of Co$_{8-x}$Fe$_x$Zn$_8$Mn$_4$, composed of randomly oriented single-crystalline grains, were synthesized by a melt growth method from pure metals of Co, Fe, Zn and Mn with nominal compositions.
Phase purity with the $\beta$-Mn-type crystal structure was confirmed using powder X-ray diffraction [Supplementary Figs. S1(a-c)]. 
Chemical compositions were examined by energy dispersive X-ray (EDX) spectroscopy, and found to be homogeneous and close to nominal compositions up to $x$ = 4.5 [Supplementary Figs. S1(d-f)]. 
Above $x$ = 5, an inhomogeneous distribution of the elements was observed.
Thus, the solubility limit of Fe is around $x$ = 4.5.
Bulk polycrystalline ingots of Co$_{8-x}$Ni$_x$Zn$_8$Mn$_4$ and Co$_{8-x}$Ru$_x$Zn$_8$Mn$_4$ were synthesized with the same method, and the solubility limit is found to be $x$ = 2 for Ni and $x$ = 2.5 for Ru, respectively.
The bulk ingots were cut into a rectangular shape with a size of $\sim$ 2 mm $\times$ 1 mm $\times$ 1 mm for magnetization, ac susceptibility and SANS measurements.
Only for Co$_{8}$Zn$_8$Mn$_4$, a single-crystalline piece grown by the Bridgman method was used.
For CBED and LTEM measurements for Co$_{8}$Zn$_8$Mn$_4$ and Co$_{3.5}$Fe$_{4.5}$Zn$_8$Mn$_4$, single-grain and single-chirality thin-plate specimens with a thickness of $\sim$ 150 nm were prepared from bulk pieces by mechanical polishing and subsequent argon ion-milling.

\subsection{Magnetization and ac magnetic susceptibility} 

Magnetization and ac magnetic susceptibility measurements were performed using the vibrating sample magnetometer (VSM) mode and ac susceptibility measurement mode, respectively, of a superconducting quantum interference device magnetometer (MPMS3, Quantum Design). 
For the ac susceptibility measurement, ac excitation field with 1 Oe and 193 Hz was applied parallel to static magnetic field. 

\subsection{SANS} 

SANS measurements were performed using the instrument SANS-I at the Paul Scherrer Institute, Switzerland.
Neutrons with a wavelength of 12 $\mathrm{\AA}$ were collimated over 18 m before the sample. The scattered neutrons were counted by a two-dimensional position-sensitive multi-detector located 20 m behind the sample. The samples were mounted on aluminum plates, and installed in a horizontal field cryomagnet. The cryomagnet was rotated (rocked) around the vertical axis in the range from $-10^\circ$ to $+10^\circ$ (1$^\circ$ step) in the magnetic field ($H$) $\perp$ beam geometry.
The SANS images shown in Supplementary Fig. S2 were obtained by summing multiple SANS measurements obtained over the above rocking scans. 
For Co$_{8}$Zn$_8$Mn$_4$, the same SANS data as our previous measurement\cite{Karube_884} were used.

\subsection{CBED and LTEM} 

For determination of the crystal chirality, CBED measurements were performed at room temperature with a transmission electron microscope (JEM-2100F) at an acceleration voltage of 200 kV. 
Convergent-beam electrons were incident along the [120] direction, and diffraction patterns containing the first-order Laue zone were obtained by a charge-coupled device camera.
For the simulation of CBED patterns, the MBFIT software\cite{Tsuda} was used.

For determination of the magnetic helicity, LTEM measurements were performed with the same specimen and the electron microscope as used in the CBED measurements. 
A magnetic field was applied perpendicular to the specimen plate and its magnitude was controlled by tuning the electric current of objective lens. 
Magnetic contrasts of skyrmions were observed as convergences (bright contrast) or divergences (dark contrast) of the electron beam on the defocused image planes. 

\subsection{Theoretical calculation}
 
Electronic structure calculations for a hypothetical model material, Co$_8$Zn$_{12}$, were performed within the generalized gradient approximation\cite{Perdew} in the framework of the density functional theory as implemented in the quantum-ESPRESSO code\cite{Giannozzi}. 
For defining the crystal geometry, the experimental structure of Co$_8$Zn$_8$Mn$_4$ was adopted. 
The DMI was calculated through spin current\cite{Kikuchi}, which was computed using interpolated electronic structures obtained by Wannier function technique\cite{Mostofi}. 
The size of the spin-orbit couplings was confirmed to be sufficiently small so that the spin current is determined by the first order of the spin-orbit coupling\cite{Freimuth_PRB2017,Koretsune_JPSJ}. 
Carrier density dependence is discussed within the rigid-band picture.

\section{Results}

\subsection{Variation of magnetic phase diagram}

First, we show the temperature ($T$) dependence of magnetization ($M$) for Co$_8$Zn$_8$Mn$_4$ ($x$ = 0), Co$_5$Fe$_3$Zn$_8$Mn$_4$ ($x$ = 3) and Co$_{3.5}$Fe$_{4.5}$Zn$_8$Mn$_4$ ($x$ = 4.5) in Figs. 2(a-c).
For $x$ = 0, $M$ exhibits a rapid increase upon cooling due to a helimagnetic transition at $T_\mathrm{c}$ $\sim$ 300 K and a gradual reduction upon cooling below $\sim$ 120 K, the latter of which corresponds to an increase in helical $q$ as explained in the next section. 
A sharp drop of $M$ around 10 K observed only in the data taken in a field-warming process after a zero-field cooling (ZFC) is due to a reentrant spin-glass transition\cite{Karube_776}. 
For $x$ = 3 with $T_\mathrm{c}$ $\sim$ 230 K, on the other hand, while a sharp change in $M$ due to a spin-glass transition is observed in the ZFC curve, the gradual decrease around 100 K is not. 
For $x$ = 4.5, the $M$ again shows a gradual reduction below $T_\mathrm{c}$ $\sim$ 135 K on cooling.
Magnetization data for all the measured samples are presented in Supplementary Fig. S3(a), showing a systematic thermal variation with the change in $x$.

In Figs. 2(d-f), the real part of ac susceptibility ($\chi^\prime$) around $T_\mathrm{c}$ is plotted against $H$ for $x$ = 0, 3 and 4.5. 
In $x$ = 0 and 4.5, the slope of $\chi^\prime$ at low fields is positive, and clear dip structures, corresponding to a SkX state, are observed near $T_\mathrm{c}$.
In $x$ = 3, on the other hand, the slope of $\chi^\prime$ at low fields is negative, and no dip structure ascribable to a SkX state is discerned at any temperature.

$H$-$T$ phase diagrams based on $\chi^\prime$ for $x$ = 0, 3 and 4.5 are shown in Fig. 2(g-i), respectively, and those for all the measured samples in Supplementary Fig. S4.
The equilibrium SkX phase near $T_\mathrm{c}$ systematically changes with increasing $x$; the SkX phase shrinks as $x$ is increased from 0 to 3, completely disappears at $x$ = 3, then reappears and expands as $x$ is further increased from 3 to 4.5.
These results are reasonably understood if the zero field helimagnetic state is suppressed in favour of a near ferromagnetic state as $x$ approaches 3.

\subsection{Determination of helical wavevectors}

To determine helical periodicity in Co$_{8-x}$Fe$_x$Zn$_8$Mn$_4$, we performed SANS measurements. 
As displayed in Supplementary Fig. S2, clear Bragg spots described by helical wavevectors $\vect{q}$ were observed in the $q_x$-$q_y$ plane below $T_\mathrm{c}$ at $H$ = 0. 
Figure 3(a) shows azimuthal-angle-averaged SANS intensity as a function of $|\vect{q}| (= q)$ for all the measured samples at 150 K (120 K for $x$ = 4.5) and 50 K (40 K for $x$ = 0). 
At 150 K (or 120 K), as $x$ is increased from 0 to 3, the Bragg peak systematically shifts to lower $q$, and then back to higher $q$ as $x$ is further increased from 3 to 4.5. 
For $x$ = 3, we observed no peak structure within the detectable $q$ range (0.01 nm$^{-1}$ $<$ $q$ $<$ 0.14 nm$^{-1}$), but only the tail of a peak centered at $q$ $\sim$ 0.
By fitting the peaks to a Gaussian function and plotting the peak center against $x$ as a helical $q$ in Fig. 3(b), we observe clearly a characteristic V-shaped curve as previously reported for Mn$_{1-x}$Fe$_x$Ge\cite{Grigoriev_MnFeGe} and Fe$_{1-x}$Co$_x$Ge\cite{Grigoriev_FeCoGe}.
On the basis of the available data points, the $q$ value is expected to be zero at $x_\mathrm{c}$ $\sim$ 2.7.
The helical periodicity $\lambda$ is also plotted against $x$ in Fig. 3(b), confirming the divergent behavior around $x_\mathrm{c}$.

As displayed in Fig. 3(a), as the temperature is lowered to 50 K the Bragg peaks shift to higher $q$ and become weaker and broader than at 150 K for all $x$, except for $x$ = 3.
The detailed temperature dependence of helical $q$ is shown in Supplementary Fig. S3(b).
A comparison between $M$ and $q$ indicates that the gradual reduction of $M$ below 120 K correlates with the increase in $q$, except for $x$ = 3.

By applying a transverse magnetic field, the positions of the Bragg spots are switched so that their wavevectors become aligned with the field direction, which corresponds to the transition from a helical multi-domain state ($\vect{q}$ $\parallel$ $\left< 100 \right>$) to a conical state ($\vect{q}$ $\parallel$ $H$) (Supplementary Fig. S2).
Note that for $x$ = 3, Bragg spots are never observed at any temperatures and any magnetic fields in the detectable $q$ range in this transverse configuration. 
These SANS results confirm that the magnetic state in $x$ = 3 is nearly ferromagnetic rather than helimagnetic, in good agreement with expectation according to the results of magnetization measurements.

The $T$-$x$ phase diagram determined from the magnetization and SANS measurements is shown in Fig. 3(c). 
The magnetic phase below $T_\mathrm{c}$ can be described as consisting of two helimagnetic regions at 0 $\leq$ $x$ $<$ $x_\mathrm{c}$ and $x_\mathrm{c}$ $<$ $x$ $\leq$ 4.5 and a nearly ferromagnetic region at $x_\mathrm{c}$ $\sim$ 2.7 in between. 
The structure of this phase diagram is consistent with the occurrence of a sign change in DMI across $x_\mathrm{c}$ as demonstrated directly in the next section.

\subsection{Determination of crystal chirality and magnetic helicity}

To determine crystal chirality and magnetic helicity, we performed CBED and LTEM measurements for thin-plate specimens with two compositions, $x$ = 0 and 4.5. 
Figures 4(a, b) and 4(e, f) show the experimentally obtained CBED patterns for $x$ = 0 and 4.5, while simulated CBED patterns for the $\beta$-Mn-type structure with $P$4$_1$32 and $P$4$_3$32 are presented in Figs. 4(c, d) and 4(g, h), respectively.
The experimental pattern for $x$ = 0 agrees well with the simulated one for $P$4$_1$32 while that for $x$ = 4.5 is in good accord with $P$4$_3$32, indicating that the crystal chirality of the examined specimens are right-handed and left-handed for $x$ = 0 and 4.5, respectively.

Next we present real-space images of skyrmions obtained by LTEM observations for exactly the same specimens as used in the CBED measurements. 
Figure 4(i)[(j)] shows the over- [under-] focused LTEM image observed for $x$ = 0 at 293 K and 45 mT, in which skyrmions with a diameter of $\sim$ 140 nm are clearly observed to form a triangular lattice. 
The observed dark (bright) dots in the over- (under-) focused image indicate that the helicity of the skyrmions is CCW consistent with the illustration shown in Fig. 1(e).

The over- [under-] focused LTEM image observed for $x$ = 4.5 at 110 K and 75 mT is displayed in Fig. 4(k)[(l)]. 
Skyrmions with a diameter of $\sim$ 130 nm are discerned and their arrangement is consistent with a finite positional disorder.
The helicity of the skyrmions is CCW, the same as $x$ = 0, as evidenced by dark (bright) dots in the over- (under-) focused image.

In Table 1, the experimentally determined crystal chirality $\Gamma_\mathrm{c}$, magnetic helicity $\gamma_\mathrm{m}$, and their correlation $\Gamma_\mathrm{c}\gamma_\mathrm{m}$ for $x$ = 0 and 4.5 are summarized.
While the crystal chiralities in the examined samples of $x$ = 0 and 4.5 are opposite to each other, the helicity of the skyrmions is the same, resulting in the opposite sign of the chirality-helicity correlation; $\Gamma_\mathrm{c}\gamma_\mathrm{m}$ = $-$1 for $x$ = 0 and $\Gamma_\mathrm{c}\gamma_\mathrm{m}$ = $+$1 for $x$ = 4.5.
Thus, it is concluded that the two helimagnetic phases shown in Fig. 3(c) possess opposite sign of $\Gamma_\mathrm{c}\gamma_\mathrm{m}$ and the sign reversal occurs at $x_\mathrm{c}$ $\sim$ 2.7.
Therefore, if the crystal chirality is fixed, the helicity of the magnetic helices and skyrmions changes sign when $x$ crosses $x_\mathrm{c}$.

\subsection{Variation of DMI}

In Fig. 5, we show the variation in DMI as a function of electron filling obtained from experiments and theoretical calculations. 
In general, the helimagnetic transition temperature $T_\mathrm{c}$ is approximately proportional to $J$ when $J$ $\gg$ $D$ like the present case, and the helical $q$ is given by $q \propto D/J$. 
Thus, we plot $\tilde{D} \equiv (-\Gamma_\mathrm{c} \gamma_\mathrm{m})T_\mathrm{c} q$ as a measure of $D$ in Fig. 5(a).
For $T_\mathrm{c}$ and $q$, values determined from the magnetization and SANS data are used, respectively.
In addition to the data for Co$_{8-x}$Fe$_x$Zn$_8$Mn$_4$, those for Co$_{8-x}$Ni$_x$Zn$_8$Mn$_4$ and Co$_{8-x}$Ru$_x$Zn$_8$Mn$_4$ are also plotted (see Supplementary Fig. S6 for $T_\mathrm{c}$ and $q$ values in Ni and Ru-doped systems).
In Co$_{8-x}$Fe$_x$Zn$_8$Mn$_4$, $\tilde{D}$ significantly decreases as $x$ is increased, or equivalently as the electron filling is decreased, and the sign reversal of $\tilde{D}$ occurs at $x_\mathrm{c}$ $\sim$ 2.7.
For both Ni and Ru doping, on the other hand, $\tilde{D}$ decreases more gradually and the sign reversal is not observed within the experimentally available concentration range ($x$ $\leq$ 2).
Although Ru doping would be identical to Fe doping in terms of band filling in the case of fully amalgamated bands of 3$d$ (Co) and 4$d$ (Ru) electron states, or would be expected to give rise to even larger effect than Fe doping due to stronger spin-orbit interaction, the experimentally observed change in $\tilde{D}$ with Ru doping is smaller than with Fe doping.
This is probably because Ru substitution makes exchange splitting smaller, as expected from the rapid suppression of $T_\mathrm{c}$ and the saturation magnetization (Supplementary Fig. S6), and hence alters the electronic structure in addition to the change in electron density and atomic spin-orbit interaction.

To evaluate the DMI theoretically, we performed \textit{ab initio} electronic structure calculations for a hypothetical model material Co$_{8}$Zn$_{12}$\cite{Xie}, where 8$c$ and 12$d$ sites are assumed to be fully occupied by Co and  Zn, respectively.
As shown in Supplementary Fig. S7, the relevant states near the Fermi level are mainly due to the Co-3$d$ band.
By using the spin-current formula described in Ref. \cite{Kikuchi}, $D$ is theoretically calculated and plotted as a function of electron band filling in Fig. 5(b).
As the carrier density falls, which corresponds to the partial substitution of Fe for Co, $D$ decreases and its sign reversal occurs at around $-4$ electron/cell.  
Upon increasing the carrier density, which corresponds to the partial substitution of Ni for Co, the sign reversal of $D$ also occurs, but at as large as $+7$ electron/cell. 
These theoretical results qualitatively reproduce the experimental results presented in Fig. 5(a). 
Therefore, the observed variation of the DMI in Co$_{8-x}$\textit{TM}$_x$Zn$_8$Mn$_4$, and its sign reversal in the case of \textit{TM} = Fe, can be understood in terms of the compositional tuning of the band filling, or equivalently in terms of the chemical potential within the complex band structure.

\section{Conclusion}

In Co$_{8-x}$Fe$_x$Zn$_8$Mn$_4$ (0 $\leq$ $x$ $\leq$ 4.5) the strength of the DMI decreases as the Fe doping proceeds, and its sign change occurs at a critical Fe concentration $x_\mathrm{c}$ $\sim$ 2.7.
Upon further Fe doping, the magnitude of the DMI increases again.
This gives rise to a reversal of the helicity of magnetic helices and skyrmions through the nearly ferromagnetic state at $x_\mathrm{c}$ under a fixed crystal chirality. 
The experimental results are well explained in terms of the change in 3$d$ band filling by the theory using a spin-current formulation and first-principles electronic structure calculation.
Band anticrossing points near the Fermi level within a complex band structure likely play an important role on these observations, as emphasized in Ref. \cite{Koretsune}.
The present study demonstrates that the DMI is controlled by the change in band filling and that in principle the sign reversal of DMI universally occurs in metallic chiral magnets with easily tunable band filling.
Our observations provide a promising route for tailoring both the size and helicity of skyrmions, and may inspire novel ideas for applications based on designed compositional gradients and that enable the local control of the DMI strength at the nanoscale.

\begin{acknowledgments}
We are grateful to A. Kikkawa for supports of sample preparation.  
This work was supported by JSPS Grant-in-Aids for Scientific Research (S) No.24224009 and for Young Scientists (B) No.17K18355, the Swiss National Science Foundation (SNSF) Sinergia network `NanoSkyrmionics (grant no. CRSII5-171003)', the SNSF projects 153451 and 166298, and the European Research Council project CONQUEST.
\end{acknowledgments}

\bibliographystyle{apsrev}

\newpage

\begin{figure}[htbp]
\begin{center}
\includegraphics[width=10cm]{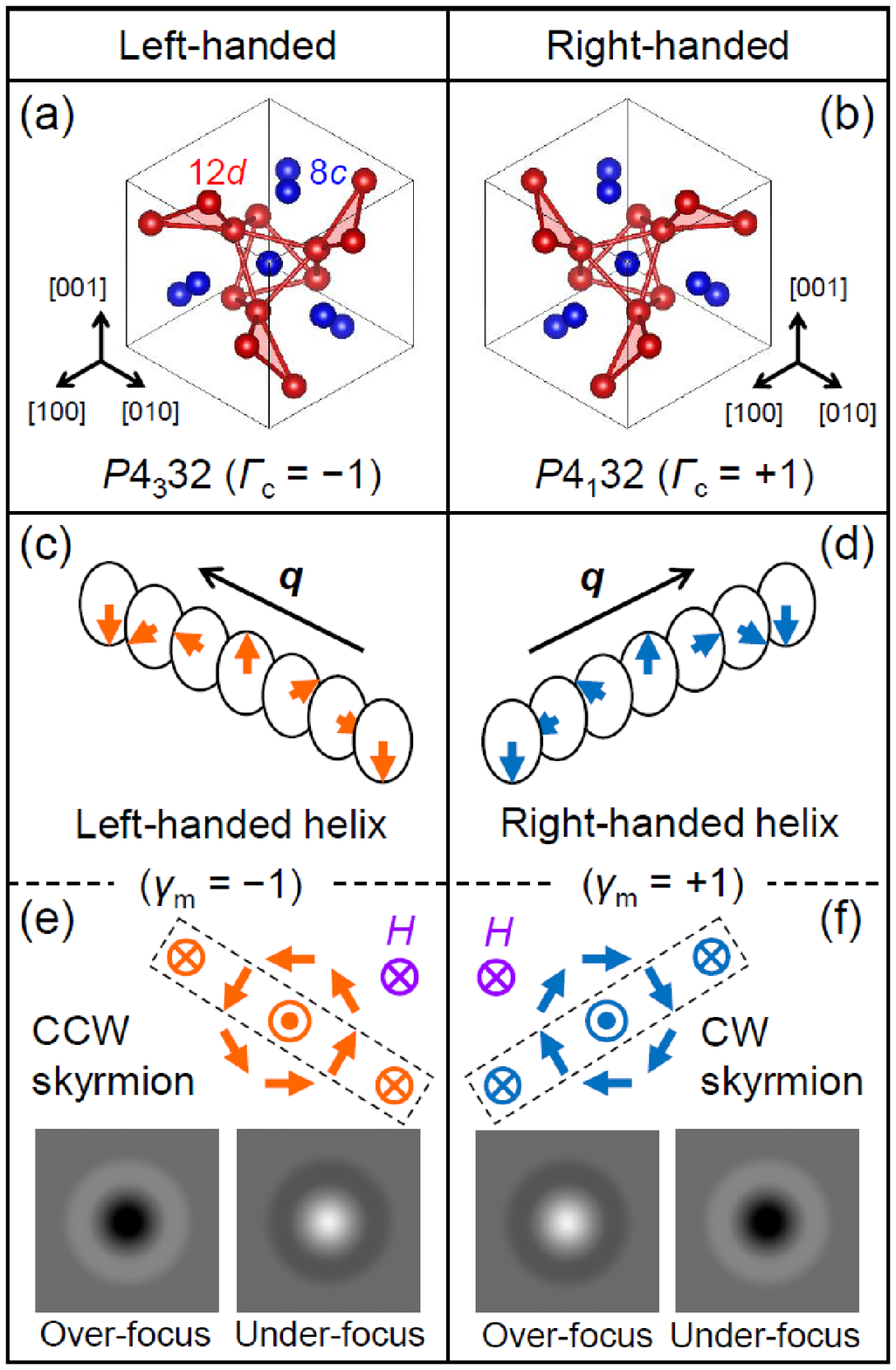}
\end{center}
\end{figure}
\noindent
FIG. 1. (a, b) Schematics of $\beta$-Mn-type chiral crystal structures as viewed along the [111] direction. 
Two enantiomers with space group $P$4$_3$32 and $P$4$_1$32 are shown. Their crystal chirality is defined as left-handed ($\Gamma_\mathrm{c}$ = $-$1) and right-handed ($\Gamma_\mathrm{c}$ = $+$1), respectively.
Blue and red circles represent 8$c$ and 12$d$ Wickoff sites, respectively.
The network of 12$d$ sites forms a windmill structure with corner-sharing triangles.
(c, d) Schematic configurations of magnetic moments for left-handed helix ($\gamma_\mathrm{m}$ = $-$1) and right-handed helix ($\gamma_\mathrm{m}$ = $+$1).
(e, f) Schematic configurations of in-plane magnetic moments of skyrmions: counterclockwise (CCW) skyrmion and clockwise (CW) skyrmion. 
Here, an external magnetic field is assumed to be applied into the page, and the magnetic moments at the periphery and the core of skyrmions are pointing into the page and out of the page, respectively.  
The magnetic configurations of the CCW and CW skyrmions along a radial direction indicated by dashed rectangles are equivalent to left-handed and right-handed helices as shown in (c) and (d), respectively.
The corresponding over-focused and under-focused Lorentz transmission electron microscope images are also displayed schematically.

\newpage

\begin{figure}[htbp]
\begin{center}
\includegraphics[width=14cm]{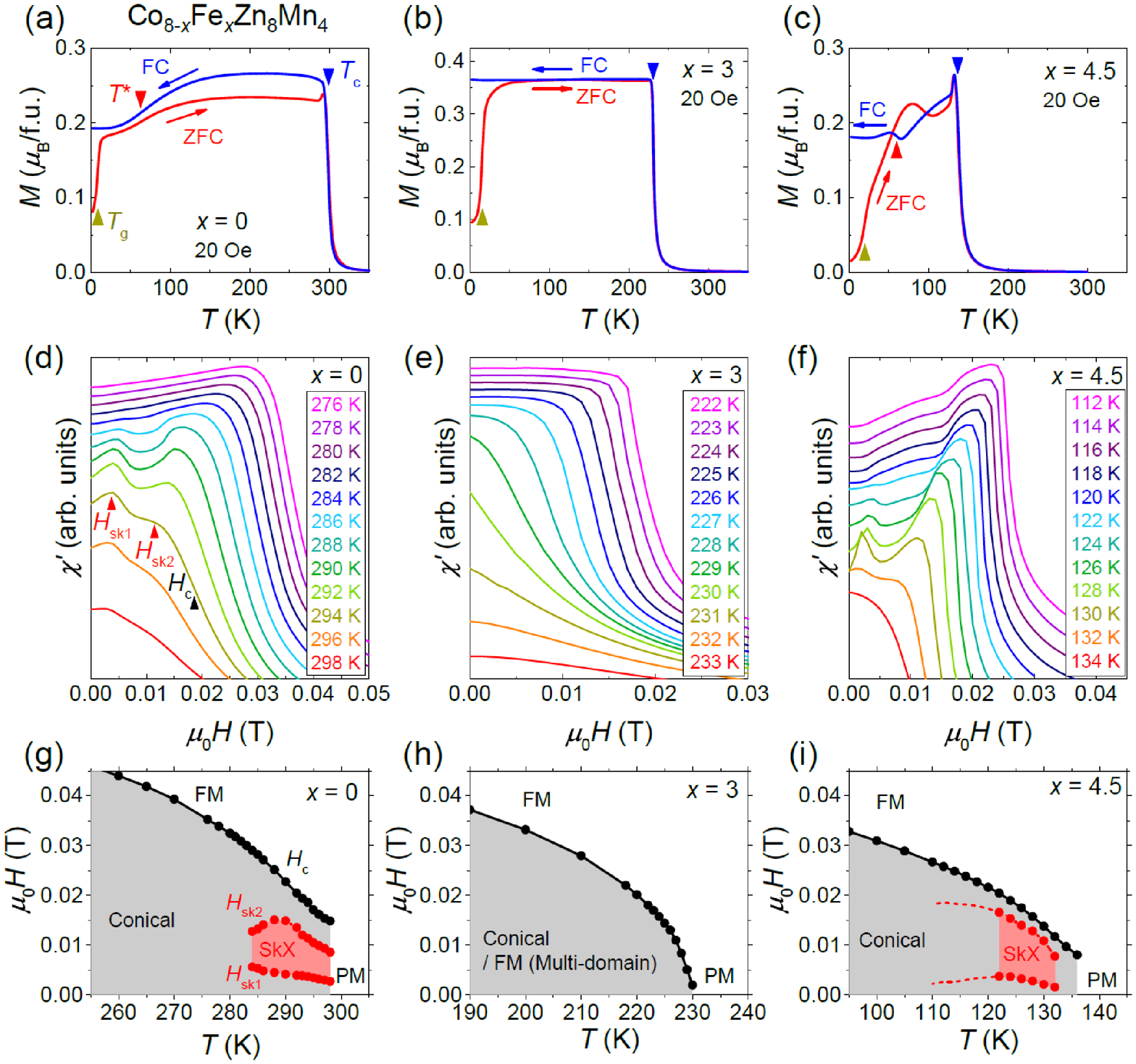}
\end{center}
\end{figure}
\noindent
FIG. 2. (a-c) Temperature ($T$) dependence of the magnetization ($M$) under a magnetic field of 20 Oe for $x$ = 0, 3 and 4.5, respectively. 
A blue line shows the data taken in a field-cooling (FC) process, and a red line the data collected in a field-warming process after zero-field cooling (ZFC) down to 2 K. 
A helimagnetic transition temperature $T_\mathrm{c}$ (blue triangle) is determined as an inflection point of the steep increase in $M$ on cooling. 
A reentrant spin-glass transition temperature $T_\mathrm{g}$ (yellow triangle) is defined as an inflection point of the sharp increase in $M$ in the warming process after ZFC. 
$T^\ast$ $\sim$ 60 K indicated with red triangles in (a) and (c) corresponds to an inflection point of gradual decrease in $M$ on cooling below 120 K, which is hardly observed for $x$ = 3 in (b).
(d-f) Magnetic field ($H$) dependence of the real part of the ac magnetic susceptibility ($\chi^\prime$) in a field-decreasing process from an induced-ferromagnetic region at several temperatures near $T_\mathrm{c}$ for $x$ = 0, 3 and 4.5, respectively. 
The phase boundaries between conical and SkX states [$H_\mathrm{sk1}$ and $H_\mathrm{sk2}$, red triangles in (d)] were determined as peak positions of $\chi^\prime$. The phase boundary between conical and induced-ferromagnetic states [$H_\mathrm{c}$, black triangle in (d)] was determined as an inflection point of $\chi^\prime$.
In $x$ = 4.5, signatures of the SkX phase become unclear below 120 K, but some anomalies are still discerned, implying a reduced volume fraction of skyrmions.
(g-i) $H$-$T$ phase diagram determined by $\chi^\prime$ for $x$ = 0, 3 and 4.5, respectively. 
In $x$ = 3, the magnetic state at low fields is either a conical state with a very long periodicity or ferromagnetic multi-domain state. 
In $x$ = 4.5, the blurred boundaries of the skyrmion phase below 120 K are plotted with dotted red lines.

\newpage

\begin{figure}[htbp]
\begin{center}
\includegraphics[width=16cm]{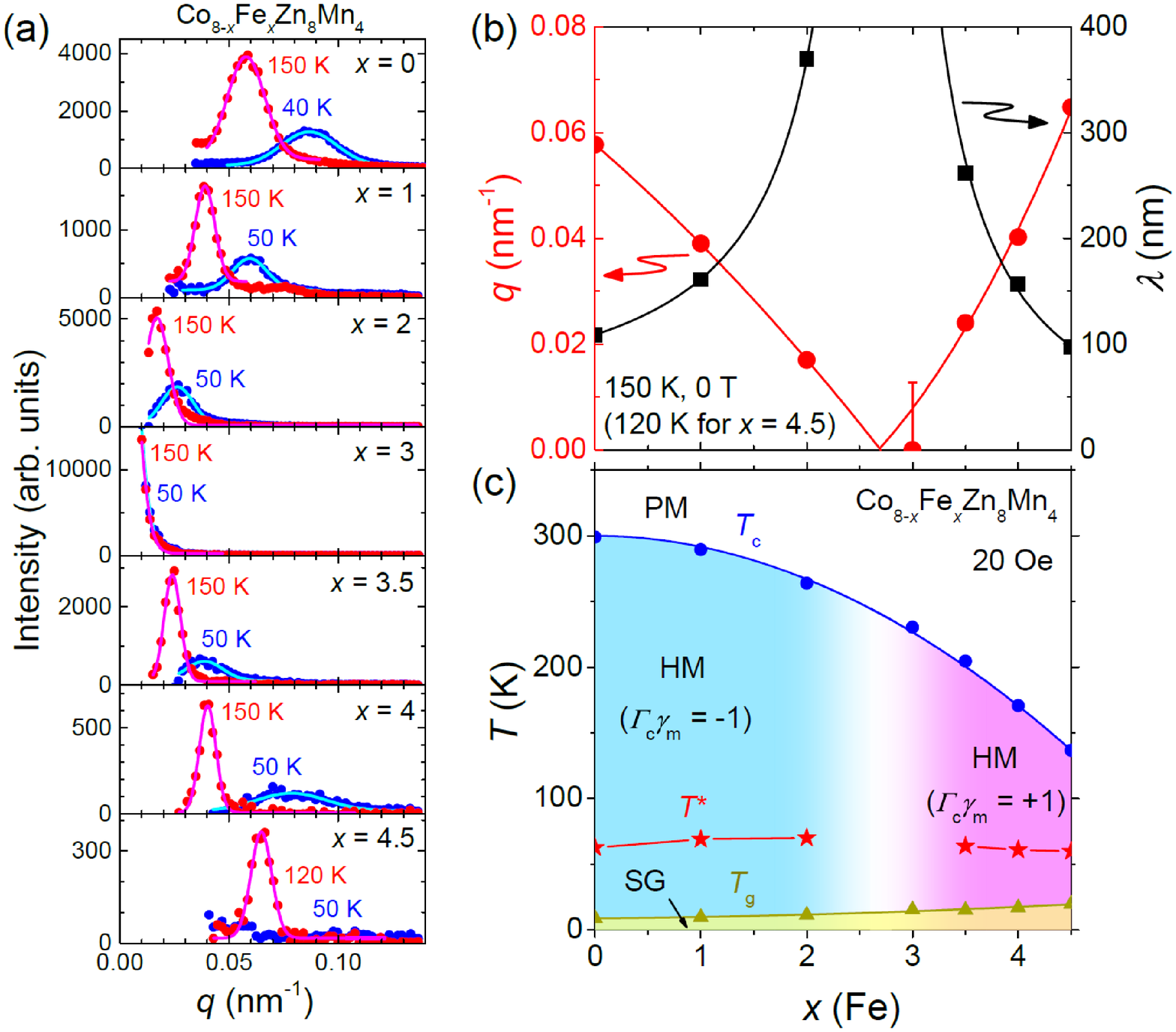}
\end{center}
\end{figure}
\noindent
FIG. 3. (a) Small-angle neutron scattering (SANS) intensity as a function of $|\vect{q}| (= q)$ for all the measured samples at 150 K (120 K for $x$ = 4.5, red circles) and 50 K (40 K for $x$ = 0, blue circles) under zero magnetic field. 
These SANS intensities are obtained by averaging over a full azimuthal-angle range of 360$^\circ$ around the origin of the two-dimensional SANS patterns (Supplementary Fig. S2). 
Data points are fitted to a Gaussian function denoted with pink lines for 150 K (120 K for $x$ = 4.5) and light-blue lines for 50 K (40 K for $x$ = 0). 
(b) $x$ dependence of helical $q$ (red circles) at 150 K (120 K for $x$ = 4.5) determined as a peak position of a Gaussian function in (a). 
For $x$ = 3, an error bar is appended as the width of the Gaussian function centered at $q$ = 0. 
By interpolating the data points (red lines), $q$ is expected to be zero at $x_\mathrm{c}$ $\sim$ 2.7. 
The helical periodicity $\lambda = 2\pi/q$ is also plotted with black squares to show the divergent behavior around $x_\mathrm{c}$.
(c) $T$-$x$ phase diagram determined by magnetization measurements at 20 Oe [see Figs. 2(a-c) and Supplementary Fig. S3(a) for details]. 
The paramagnetic (PM) - helimagnetic (HM) transition temperature $T_\mathrm{c}$ is shown by blue circles, and the spin glass (SG) transition temperature $T_\mathrm{g}$ is denoted by yellow triangles. 
$T^\ast$ (red asterisks) indicates a temperature corresponding to the inflection point during a gradual reduction in $M$ on cooling, and also to an increase in helical $q$ (see Supplementary Fig. S3).
Two helimagnetic phases (light-blue and pink areas) separated by a nearly ferromagnetic region (white area below $T_\mathrm{c}$) are characterized by opposite signs of $\Gamma_\mathrm{c}\gamma_\mathrm{m}$.

\newpage

\begin{figure}[htbp]
\begin{center}
\includegraphics[width=16cm]{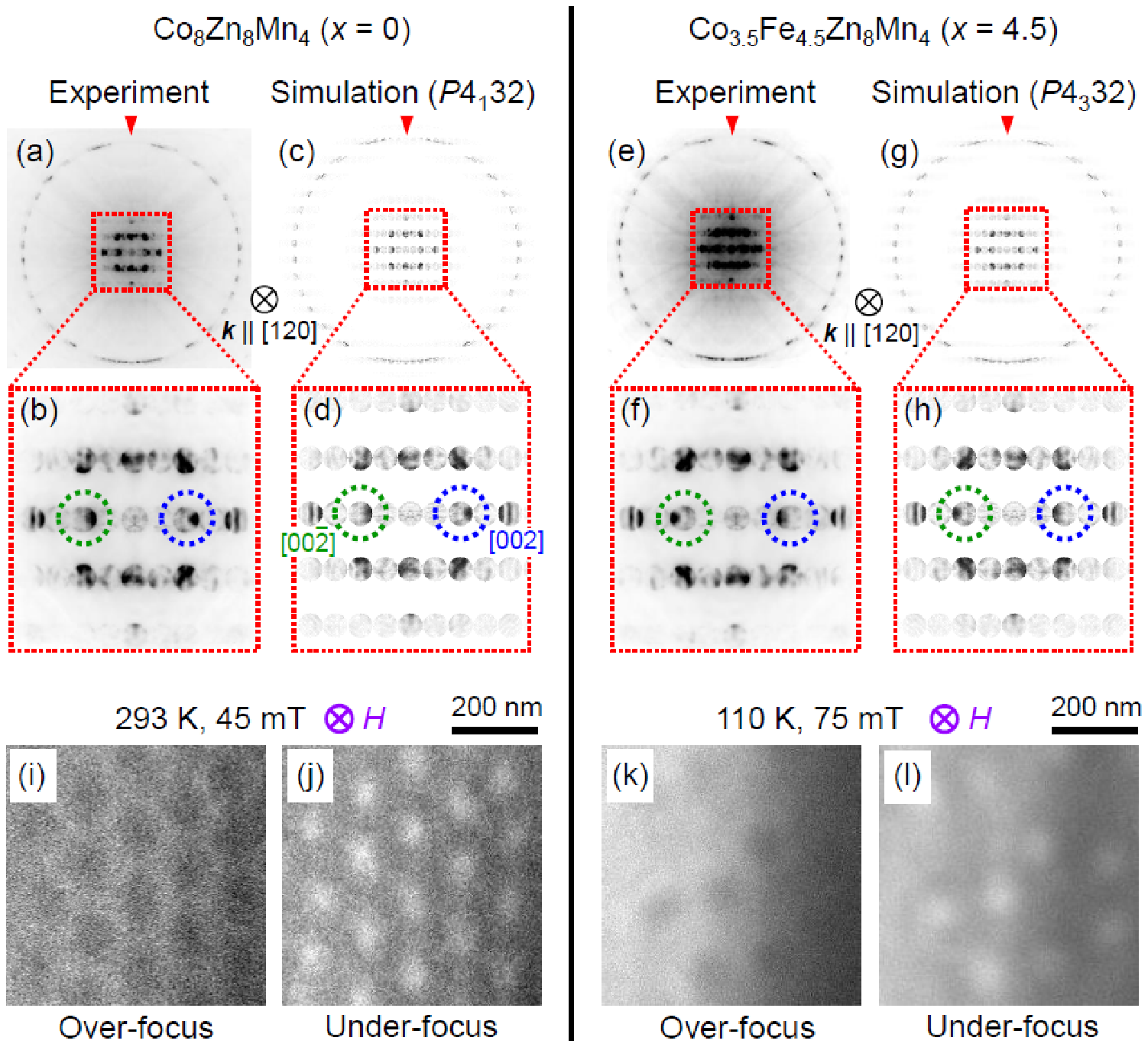}
\end{center}
\end{figure}
\noindent
FIG. 4. (a, b) Convergent-beam electron diffraction (CBED) patterns observed from a thin-plate specimen of Co$_8$Zn$_8$Mn$_4$.
(c, d) Simulated CBED patterns for right-handed crystal with the space group of $P$4$_1$32.
(e, f) CBED patterns observed in a thin-plate specimen of Co$_{3.5}$Fe$_{4.5}$Zn$_8$Mn$_4$.
(g, h) Simulated CBED patterns for left-handed crystal with the space group of $P$4$_3$32.
For all the panels (a-h), the incident electron beam direction is along the [120] direction.
A red triangle in upper panels (a, c, e, g) indicates the disconnected intensity at the top part of circular diffraction that makes a distinction with the connected intensity at the bottom part, reflecting broken inversion symmetry.
Lower panels (b, d, f, h) are expanded images for the red dotted-line enclosed areas of (a, c, e, g), respectively.
Blue and green dotted circles in (b, d, f, h) highlight [002] and [00$\overline{2}$] diffraction signals, respectively. 
(i-l) Lorentz transmission electron microscope (LTEM) images observed for the same thin-plate specimens as used in the CBED measurements. 
Panel (i)[(j)] is an over- [under-] focused LTEM image for Co$_8$Zn$_8$Mn$_4$, measured at 293 K and under a magnetic field of 45 mT applied perpendicular (into the page) to the thin specimen. 
Panel (k)[(l)] is an over- [under-] focused LTEM image for Co$_{3.5}$Fe$_{4.5}$Zn$_8$Mn$_4$, measured at 110 K and 75 mT. 

\newpage

\begin{table}[htbp]
\begin{center}
\begin{tabular}{cc cc} \hline
\:\:\:\: \:\:\:\: & \:\:\:\: $\Gamma_\mathrm{c}$ \:\:\:\: & \:\:\:\: $\gamma_\mathrm{m}$ \:\:\:\: & \:\:\:\: $\Gamma_\mathrm{c} \gamma_\mathrm{m} \:\:\:\: $ \\ \hline 
Co$_8$Zn$_8$Mn$_4$ & $+1$ & $-1$ & $-1$ \\ 
Co$_{3.5}$Fe$_{4.5}$Zn$_8$Mn$_4$ & $-1$ & $-1$ & $+1$ \\ \hline
\end{tabular}
\end{center}
\end{table}
\noindent
TABLE. 1. Crystal chirality $\Gamma_\mathrm{c}$, magnetic helicity $\gamma_\mathrm{m}$ and their correlation $\Gamma_\mathrm{c}\gamma_\mathrm{m}$ for Co$_8$Zn$_8$Mn$_4$ and Co$_{3.5}$Fe$_{4.5}$Zn$_8$Mn$_4$ determined by CBED and LTEM measurements.

\newpage

\begin{figure}[htbp]
\begin{center}
\includegraphics[width=16cm]{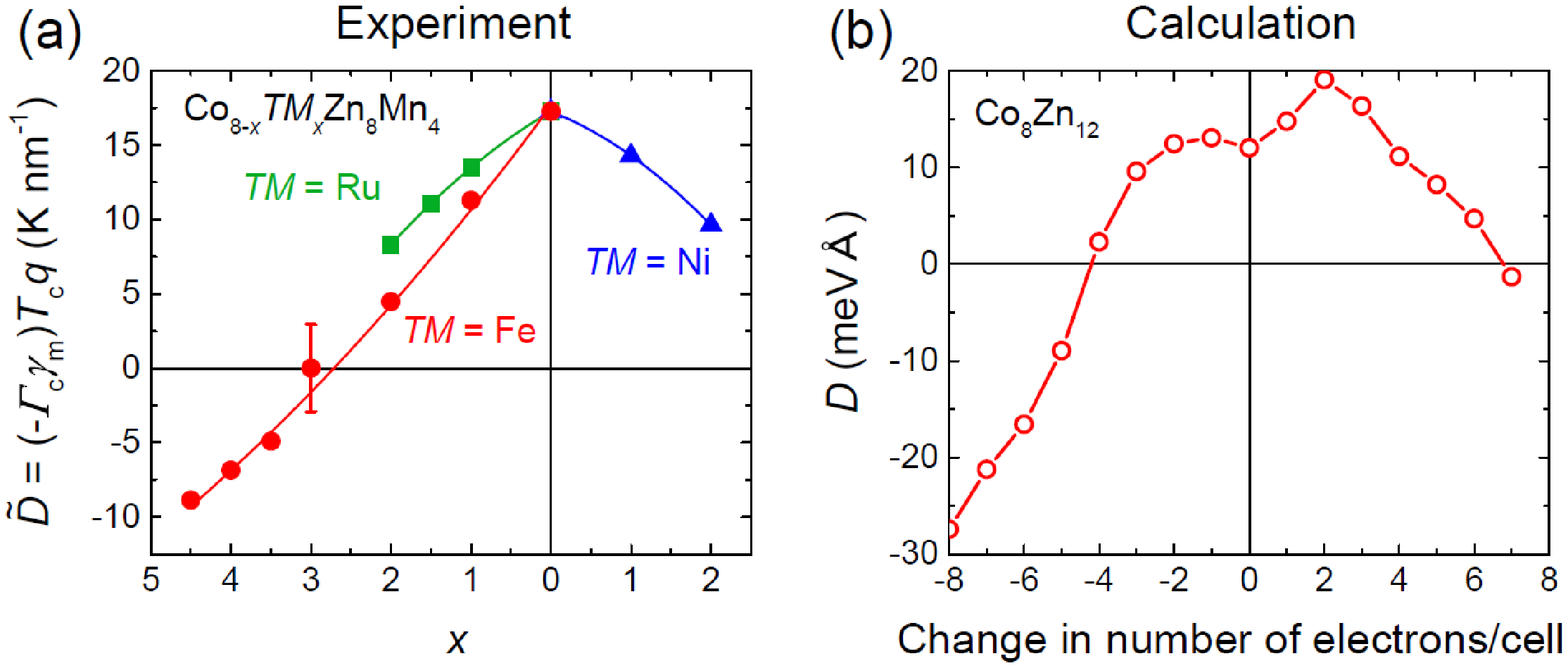}
\end{center}
\end{figure}
\noindent
FIG. 5. 
(a) Dopant concentration ($x$) dependence of $\tilde{D}$ $\equiv$ $(-\Gamma_\mathrm{c}\gamma_\mathrm{m})T_\mathrm{c}q$, which is proportional to Dzyaloshinskii-Moriya interaction $D$, in Co$_{8-x}$\textit{TM}$_x$Zn$_8$Mn$_4$.
Here, \textit{TM} = Fe (red circles) and Ru (green squares) correspond to hole doping while \textit{TM} = Ni (blue triangles) to electron doping. 
The values of $T_\mathrm{c}$ and $q$ are determined by magnetization and SANS measurements, respectively. 
See Supplementary Figs. S5 and S6 for details of magnetization and SANS measurements for \textit{TM} = Ni and Ru. 
The minus sign is appended to the definition of $\tilde{D}$ due to the following reason; both the experiment and theory consistently show that the right-handed crystal ($\Gamma_\mathrm{c}$ = $+$1) of the Fe non-doped material exhibits a left-handed helix ($\gamma_\mathrm{m}$ = $-$1). In our theoretical definition, this corresponds to $D$ $>$ 0, therefore we included the minus sign to keep the consistency.
(b) Theoretically calculated value of $D$ plotted as a function of electron number per unit cell, which is taken as the origin for a hypothetical model compound Co$_8$Zn$_{12}$, where 8$c$ and 12$d$ sites are occupied by only Co and Zn, respectively.
The calculated band structure and density of states are shown in Supplementary Fig. S7.

\end{document}


\title{Supplementary Material for \\ ``Controlling the helicity of magnetic skyrmions in a $\beta$-Mn-type high-temperature chiral magnet"}
\author{K.~Karube}
\affiliation{RIKEN Center for Emergent Matter Science (CEMS), Wako 351-0198, Japan.}
\author{K.~Shibata}
\affiliation{RIKEN Center for Emergent Matter Science (CEMS), Wako 351-0198, Japan.}
\author{J.~S.~White}
\affiliation{Laboratory for Neutron Scattering and Imaging (LNS), Paul Scherrer Institute (PSI),
CH-5232 Villigen, Switzerland.}
\author{T.~Koretsune}
\affiliation{Department of Physics, Tohoku University, Sendai 980-8578, Japan.}
\author{X.~Z.~Yu}
\affiliation{RIKEN Center for Emergent Matter Science (CEMS), Wako 351-0198, Japan.}
\author{Y.~Tokunaga}
\affiliation{Department of Advanced Materials Science, University of Tokyo, Kashiwa 277-8561, Japan.}
\author{H.~M.~R\o nnow}
\affiliation{Laboratory for Quantum Magnetism (LQM), Institute of Physics, \'Ecole Polytechnique F\'ed\'erale de Lausanne (EPFL), CH-1015 Lausanne,
Switzerland.}
\author{R.~Arita}
\affiliation{RIKEN Center for Emergent Matter Science (CEMS), Wako 351-0198, Japan.}
\affiliation{Department of Applied Physics, University of Tokyo, Bunkyo-ku 113-8656, Japan.}
\author{T.~Arima}
\affiliation{RIKEN Center for Emergent Matter Science (CEMS), Wako 351-0198, Japan.}
\affiliation{Department of Advanced Materials Science, University of Tokyo, Kashiwa 277-8561, Japan.}
\author{Y.~Tokura}
\affiliation{RIKEN Center for Emergent Matter Science (CEMS), Wako 351-0198, Japan.}
\affiliation{Department of Applied Physics, University of Tokyo, Bunkyo-ku 113-8656, Japan.}
\author{Y.~Taguchi}
\affiliation{RIKEN Center for Emergent Matter Science (CEMS), Wako 351-0198, Japan.}

\maketitle

\newpage

\subsection{Sample growth and characterization}

Bulk polycrystalline samples of Co$_{8-x}$Fe$_x$Zn$_8$Mn$_4$, composed of randomly oriented single-crystalline grains, were synthesized by a melt growth method; pure metals of Co, Fe, Zn and Mn with nominal compositions were sealed in an evacuated quartz tube, heated up to 1200$^\circ$C, slowly cooled down to 900$^\circ$C, annealed for 1 day, and slowly cooled down to room temperature.
Phase purity with the $\beta$-Mn-type structure was confirmed using X-ray diffraction (XRD) on a powder obtained by crushing an ingot for each sample.
The obtained powder XRD profiles for $x$ = 3 and 4.5 are shown in Figs. S1(a) and (b), respectively.
The lattice constant of the $\beta$-Mn-type cubic structure, determined by Rietveld analysis, increases as $x$ is increased, and saturates around $x$ $\sim$ 4 [Fig. S1(c)].
Chemical compositions were examined by energy dispersive X-ray spectroscopy (EDX) and found to be close to nominal ones up to $x$ = 4.5 [Fig. S1(d)].
Two-dimensional mappings in terms of EDX confirmed homogeneous distributions of the 4 elements over a length scale of both a 100-$\mu$m order [Fig. S1(e)] and a 100-nm order [Fig. S1(f)], comparable to skyrmion diameter, in the sample of $x$ = 4.5.
Since we observed macroscopic phase separation in the samples with $x$ $\geq$ 5, the solubility limit of Fe in Co$_{8-x}$Fe$_x$Zn$_8$Mn$_4$ is around $x$ = 4.5 (22.5\%).
This value is comparable with those reported in literatures; The solubility limit of Fe in Co$_{10}$Zn$_{10}$ is approximately 10\%\cite{Takayama} and that in $\beta$-Mn is approximately 30\%\cite{Okamoto}.

Bulk polycrystalline ingots of Co$_{8-x}$Ni$_x$Zn$_8$Mn$_4$ and Co$_{8-x}$Ru$_x$Zn$_8$Mn$_4$ were synthesized with the same method.
The maximum miscible amount is $x$ = 2 (10\%) for Ni and $x$ = 2.5 (12.5\%) for Ru, respectively.

Bulk ingots were cut to a rectangular shape with a size of $\sim$ 2 mm $\times$ 1 mm $\times$ 1 mm for magnetization, ac magnetic susceptibility and small-angle neutron scattering (SANS) measurements.

\subsection{SANS measurements}

Figure S2 shows SANS patterns on two-dimensional $q_x$-$q_y$ planes observed in Co$_{8-x}$Fe$_x$Zn$_8$Mn$_4$ (1 $\leq$ $x$ $\leq$ 4.5) at 150 K (120 K for $x$ = 4.5).
SANS patterns observed at zero field are not ring-like patterns expected for polycrystalline samples, but spots expected for single-crystalline samples due to large grain size of the samples.
Bragg spots are observed irregularly in terms of azimuthal-angle position and the number (2 or 4), depending on the compositions.
This is because helical-$\vect{q}$ domains also appear to be large, and preferred helical-$\vect{q}$ orientation is along $<$100$>$ directions in this system, but the orientation of crystalline axes was not aligned to horizontal nor vertical direction in this measurement.
SANS intensity integrated over full azimuthal-angle region is shown as a function of $|\vect{q}| (= q)$ in Fig. 3(a) in the main text.
When a transverse magnetic field ($H$) is applied, SANS patterns change to 2 horizontal spots, which corresponds to the transition from a helical multi-domain state ($\vect{q}$ $\parallel$ $<$100$>$) to a conical state ($\vect{q}$ $\parallel$ $H$). 
Bragg spots both for helical and conical states shift to lower $q$ region as $x$ is increased from 1 to 3, and to higher $q$ region as $x$ is further increased from 3 to 4.5. 
At $x$ = 3, no Bragg spots are observed either at zero field [Fig. S2(e)] or under a transverse field [Fig. S2(f)] in the detectable $q$ range (0.01 nm$^{-1}$ $<$ $q$ $<$ 0.14 nm$^{-1}$).
This systematic variation in SANS patterns with $x$ confirms that a helical state transforms to a nearly ferromagnetic state as $x$ = 3 is approached.

\subsection{Temperature dependence of magnetization and helical wavevector}

Temperature ($T$) dependence of magnetization ($M$) and helical wavevector ($q$) for Co$_{8-x}$Fe$_x$Zn$_8$Mn$_4$ (0 $\leq$ $x$ $\leq$ 4.5) is shown in Figs. S3(a) and (b), respectively. 
By comparing $M$ and $q$, a gradual reduction in $M$ below 120 K on cooling appears to correspond to an increase in $q$. 
As $x$ = 3 is approached, the gradual reduction in $M$ becomes weaker and completely disappears for $x$ = 3, which is in good accord with nearly zero $q$-value, i.e. nearly ferromagnetic, in the whole temperature range for $x$ = 3. 
Reentrant spin glass transitions, as identified by the rapidly reduced $M$ below $\sim$ 10 K observed only in a field-warming process after zero-field cooling (ZFC), are discernible in all the compositions, including $x$ = 3, and thus independent of the ordered states, i.e. helimagnetic or ferromagnetic, at higher temperatures.

\subsection{Ac magnetic susceptibility and skyrmion phase diagram}

Figures S4(a-n) show magnetic field ($H$) dependence of the real part ($\chi^\prime$) and the imaginary part ($\chi^{\prime\prime}$) of the ac magnetic susceptibility near $T_\mathrm{c}$ in Co$_{8-x}$Fe$_x$Zn$_8$Mn$_4$ (0 $\leq$ $x$ $\leq$ 4.5).
For $x$ = 0, 1, 4 and 4.5, $\chi^\prime$ shows clear dip structures and $\chi^{\prime\prime}$ exhibits double peaks, hallmarks of a skyrmion crystal (SkX) state.
For $x$ = 2 and 3.5, on the other hand, the dip structure is not discerned in $\chi^\prime$ while $\chi^{\prime\prime}$ still shows weak double peaks in a narrow temperature range, indicating that skyrmions barely form.
For $x$ = 3, neither the dip structure in $\chi^{\prime}$ nor the double peaks in $\chi^{\prime\prime}$ are discerned, suggesting that skyrmions do not form any more.

$H$-$T$ phase diagrams near $T_\mathrm{c}$ determined by the ac magnetic susceptibility data are shown in Figs. S4(o-u).
The equilibrium SkX phase varies systematically with $x$; The SkX phase shrinks as $x$ is increased from 0 to 2, completely disappears at $x$ = 3, reappears and expands as $x$ is changed from 3.5 to 4.5.
This result also indicates transformation from a helical state (with SkX phase in magnetic fields) to a nearly ferromagnetic state around $x$ = 3.

\subsection{Ni and Ru dopings}

Figure S5 displays magnetization and SANS results in Co$_{8-x}$Ni$_x$Zn$_8$Mn$_4$ (0 $\leq$ $x$ $\leq$ 2) and Co$_{8-x}$Ru$_x$Zn$_8$Mn$_4$ (0 $\leq$ $x$ $\leq$ 2). 
Temperature dependence of $M$ is shown in Fig. S5(a).
While $T_\mathrm{c}$ decreases with increasing $x$, a gradual reduction in $M$ below 120 K is similarly observed in all the compositions for Ni and Ru dopings. 
We also confirmed that the temperature window of an equilibrium SkX phase just below $T_\mathrm{c}$, determined by ac magnetic susceptibility measurements, is almost independent of $x$ both for Ni and Ru dopings.

Figure S5(b) presents azimuthally-integrated SANS intensity versus $q$ measured at 150 K (120 K for $x$ = 2 in Ru doping) and 50 K under zero field.
The observed $q$ positions for Bragg peaks are almost independent of $x$ both for Ni and Ru dopings. 

Temperature dependence of helical $q$, determined as a peak center of the Gaussian function fit to the intensity, is shown in Fig. S5(c).
The increase in helical $q$ below 120 K observed in all the compositions for Ni and Ru dopings shows good correspondence to the gradual reduction in $M$ shown in Fig. S5(a).
Based on these results, it is concluded that Ni and Ru dopings within the range of 0 $\leq$ $x$ $\leq$ 2 do not induce a transformation from helimagnetic state to ferromagnetic state as observed in the case of Fe doping.

Experimentally-obtained important physical parameters for Co$_{8-x}$\textit{TM}$_x$Zn$_8$Mn$_4$ (\textit{TM} = Fe, Ni and Ru) are summarized in Fig. S6.
Ni and Ru dopings rapidly suppress $T_\mathrm{c}$ that represents ferromagnetic exchange interaction $J$, and saturation magnetization $M_\mathrm{s}$ that reflects exchange splitting of the bands while the suppression in the case of Fe doping is more gradual [Figs. S6(a) and (b)].
On the other hand, helical $q$ changes little with Ni and Ru dopings in contrast to significant change in $q$ in the case of Fe doping [Fig. S6(c)].
As a result, $(-\Gamma_\mathrm{c}\gamma_\mathrm{m})T_\mathrm{c}q$ that represents Dzyaloshinskii-Moriya interaction $D$ decreases with Ni and Ru dopings but a little bit more slowly than the case of Fe doping.

\subsection{\textit{Ab initio} electronic structure calculation for Co$_8$Zn$_{12}$}

Figure S7 shows the electronic structure of Co$_8$Zn$_{12}$ calculated by means of \textit{ab initio} density-functional theory.
Co$_8$Zn$_{12}$ is a hypothetical material with the $\beta$-Mn-type chiral structure\cite{Xie}, where 8$c$ and 12$d$ Wyckoff sites are assumed to be fully occupied by 8 Co atoms and 12 Zn atoms, respectively.
For a lattice constant, experimentally obtained value (6.374 $\mathrm{\AA}$) for Co$_8$Zn$_8$Mn$_4$ was used. 
The obtained band structure is shown in Fig. S7(a).
The relevant bands close to Fermi level from $-$4 eV to 2 eV consist of mainly Co-3$d$ states while Zn states lie deeply below $-$6 eV.
In Fig. S7(b), the detailed band structure near the Fermi level is displayed with colors representing the weight of spin component. 
The density of states (DOS) for up-spin and down-spin states are also plotted in Fig. S7(c).

\bibliographystyle{apsrev}

\newpage

\begin{figure}[htbp]
\begin{center}
\includegraphics[width=16cm]{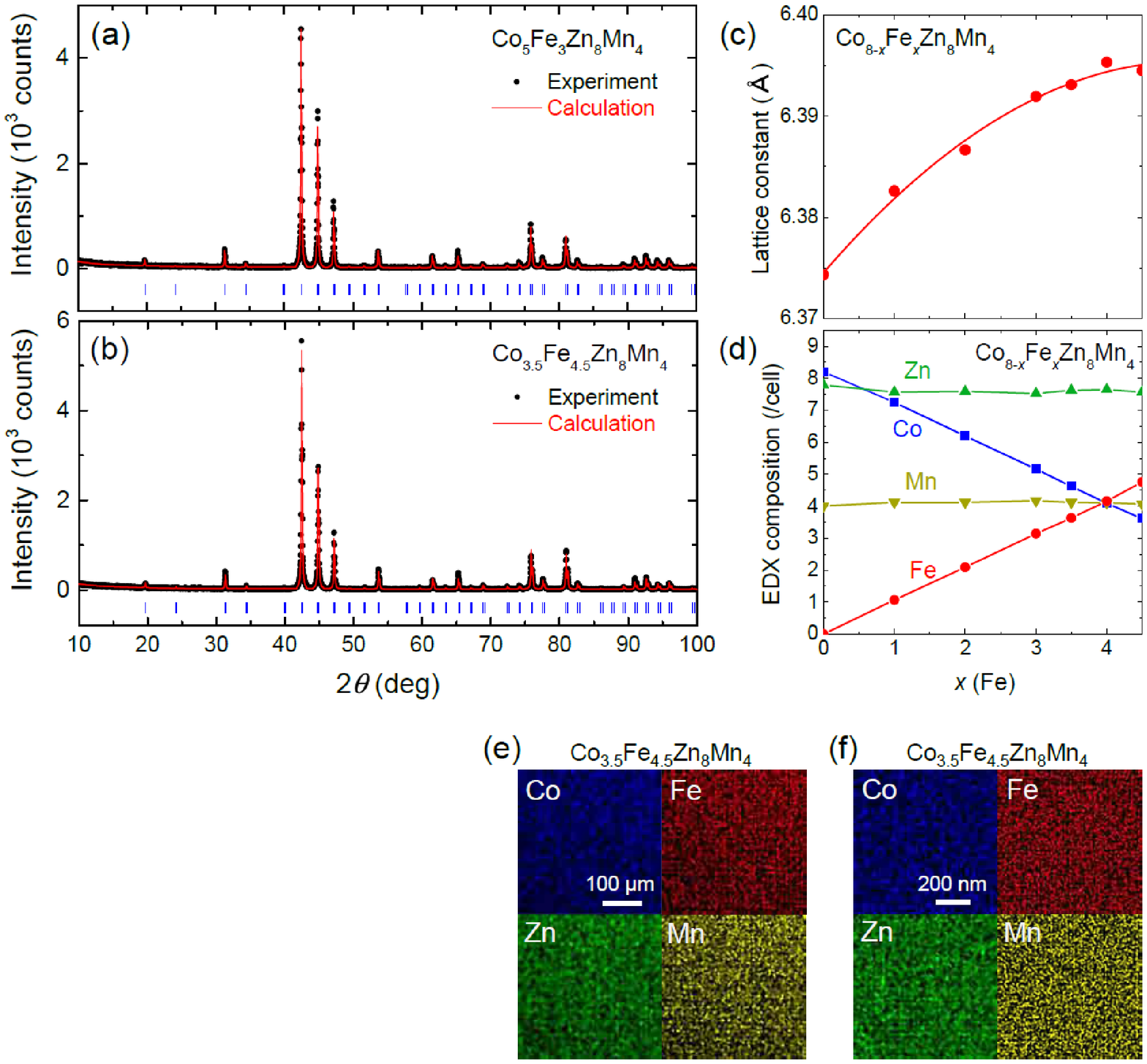}
\end{center}
\end{figure}
\noindent
FIG. S1. (a, b) Powder X-ray diffraction (XRD) profiles for (a) Co$_{5}$Fe$_{3}$Zn$_8$Mn$_4$ and (b) Co$_{3.5}$Fe$_{4.5}$Zn$_8$Mn$_4$ measured at room temperature. 
Experimental data (black circles) are plotted together with Rietveld refinements (red line) with a $\beta$-Mn-type structure. 
Peak positions are indicated by blue bars. 
(c) $x$ dependence of lattice constant of the $\beta$-Mn-type cubic structure in Co$_{8-x}$Fe$_x$Zn$_8$Mn$_4$ determined by Rietveld analyses for the powder XRD data.
(d) $x$ dependence of actual concentration (atom numbers per unit cell) of Co, Fe, Zn and Mn, determined by energy dispersive X-ray (EDX) spectroscopy.
(e, f) Two-dimensional mappings of constituent elements, Co, Fe, Zn and Mn, in terms of EDX for Co$_{3.5}$Fe$_{4.5}$Zn$_8$Mn$_4$ over a length scale of (e) several hundreds of $\mu$m and (f) several hundreds of nm.

\newpage

\begin{figure}[htbp]
\begin{center}
\includegraphics[width=12cm]{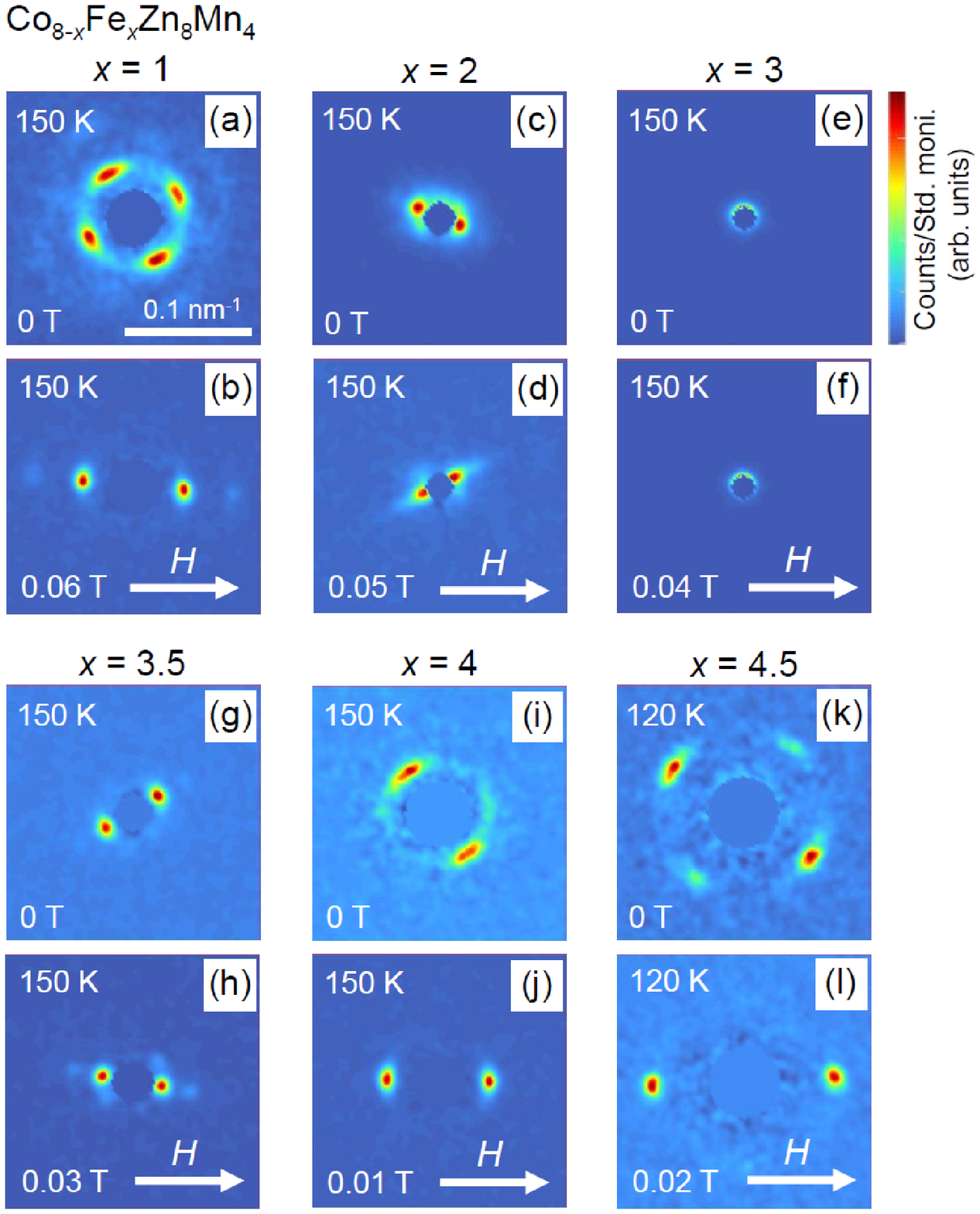}
\end{center}
\end{figure}
\noindent
FIG. S2. Small-angle neutron scattering (SANS) patterns in Co$_{8-x}$Fe$_x$Zn$_8$Mn$_4$. (a, b) $x$ = 1 at 150 K under zero field and a transverse field of 0.06 T. (c, d) $x$ = 2 at 150 K under zero field and a transverse field of 0.05 T. (e, f) $x$ = 3 at 150 K under zero field and a transverse field of 0.04 T. (g, h) $x$ = 3.5 at 150 K under zero field and a transverse field of 0.03 T. (l, j) $x$ = 4 at 150 K under zero field and a transverse field of 0.01 T. (k, l) $x$ = 4.5 at 120 K under zero field and a transverse field of 0.02 T. 
For all the panels, the scales of horizontal axis and vertical axis are fixed to be $-$0.1 nm$^{-1}$ $\leq$ $q_x$ $\leq$ 0.1 nm$^{-1}$ and $-$0.1 nm$^{-1}$ $\leq$ $q_y$ $\leq$ 0.1 nm$^{-1}$, respectively. 
The intensity scale of the color plots varies between each panel. 
Central region of each panel ($q$ $\sim$ 0) is masked to avoid the direct beam.

\newpage

\begin{figure}[htbp]
\begin{center}
\includegraphics[width=12cm]{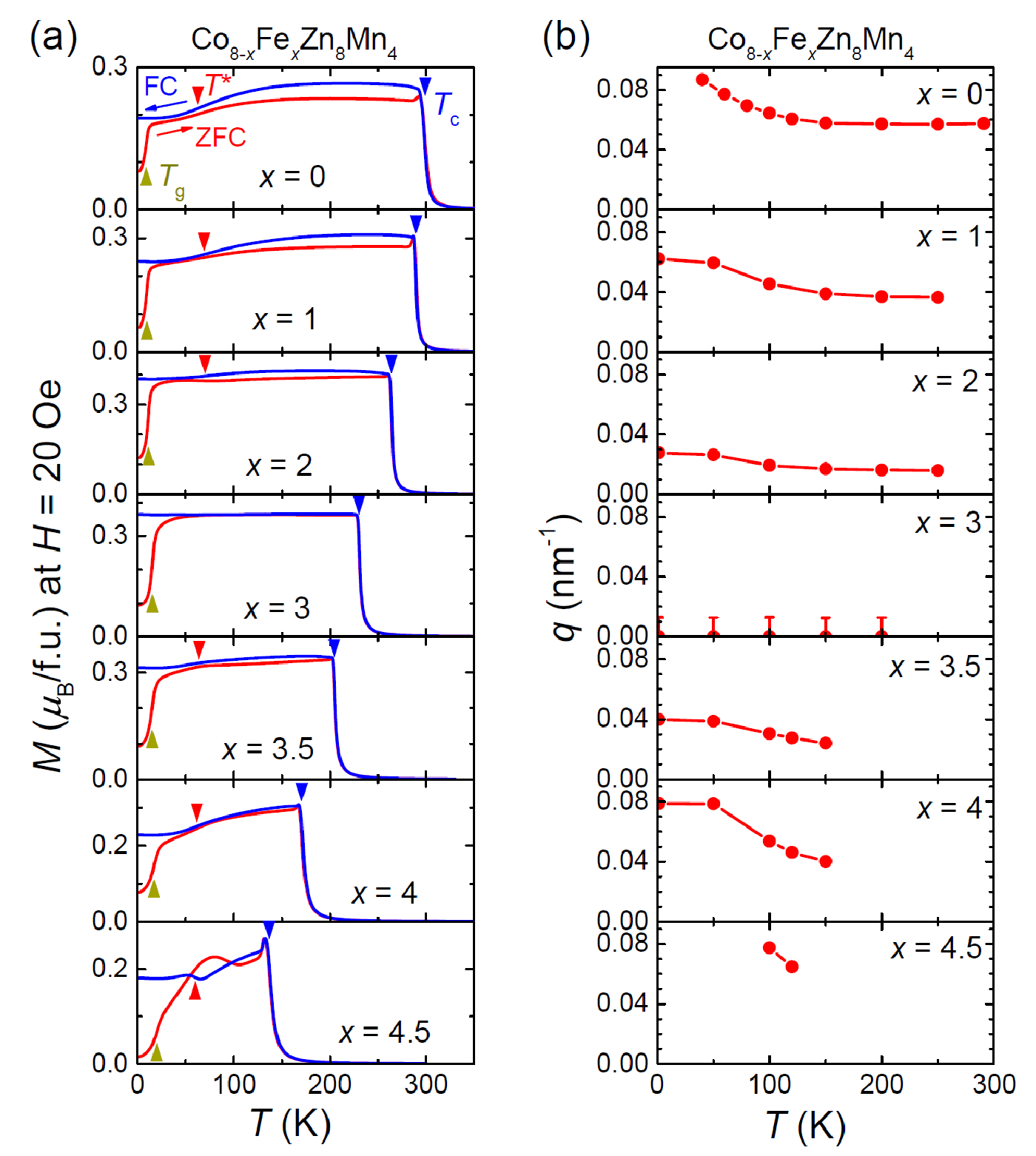}
\end{center}
\end{figure}
\noindent
FIG. S3. (a) Temperature ($T$) dependence of magnetization ($M$) under a magnetic field of 20 Oe in Co$_{8-x}$Fe$_x$Zn$_8$Mn$_4$. 
A blue line shows the data collected in a field-cooling (FC) process, and a red line represents the data taken in a field-warming process after a zero-field cooling (ZFC) down to 2 K. 
A helimagnetic transition temperature $T_\mathrm{c}$ (blue triangle) was determined as an inflection point of sharp increase in $M$ on cooling. 
A reentrant spin-glass transition temperature $T_\mathrm{g}$ (yellow triangle) was determined as an inflection point of sharp increase in $M$ in the field-warming process after ZFC. 
Red triangles around $T^\ast$ $\sim$ 60 K show an inflection point of gradual decrease in $M$ upon cooling below 120 K.
These characteristic temperatures are plotted in the $T$-$x$ phase diagram in Fig. 3(c) in the main text.
(b) Temperature dependence of helical wavevector $q$ in Co$_{8-x}$Fe$_x$Zn$_8$Mn$_4$.
The helical $q$ value is determined as the peak center of Gaussian function fitted to azimuthally-integrated SANS intensity as a function of $|\vect{q}|$ at zero field, which is shown in Fig 3(a) in the main text.
For $x$ = 3, error bars are appended as the width of Gaussian function centered at $q$ = 0. 
For $x$ = 4.5, at 50 K and 1.5 K clear Bragg peaks are not observed probably due to weakened intensities and/or large $q$ values beyond the detectable $q$ range.


\begin{figure}[htbp]
\begin{center}
\includegraphics[width=17cm]{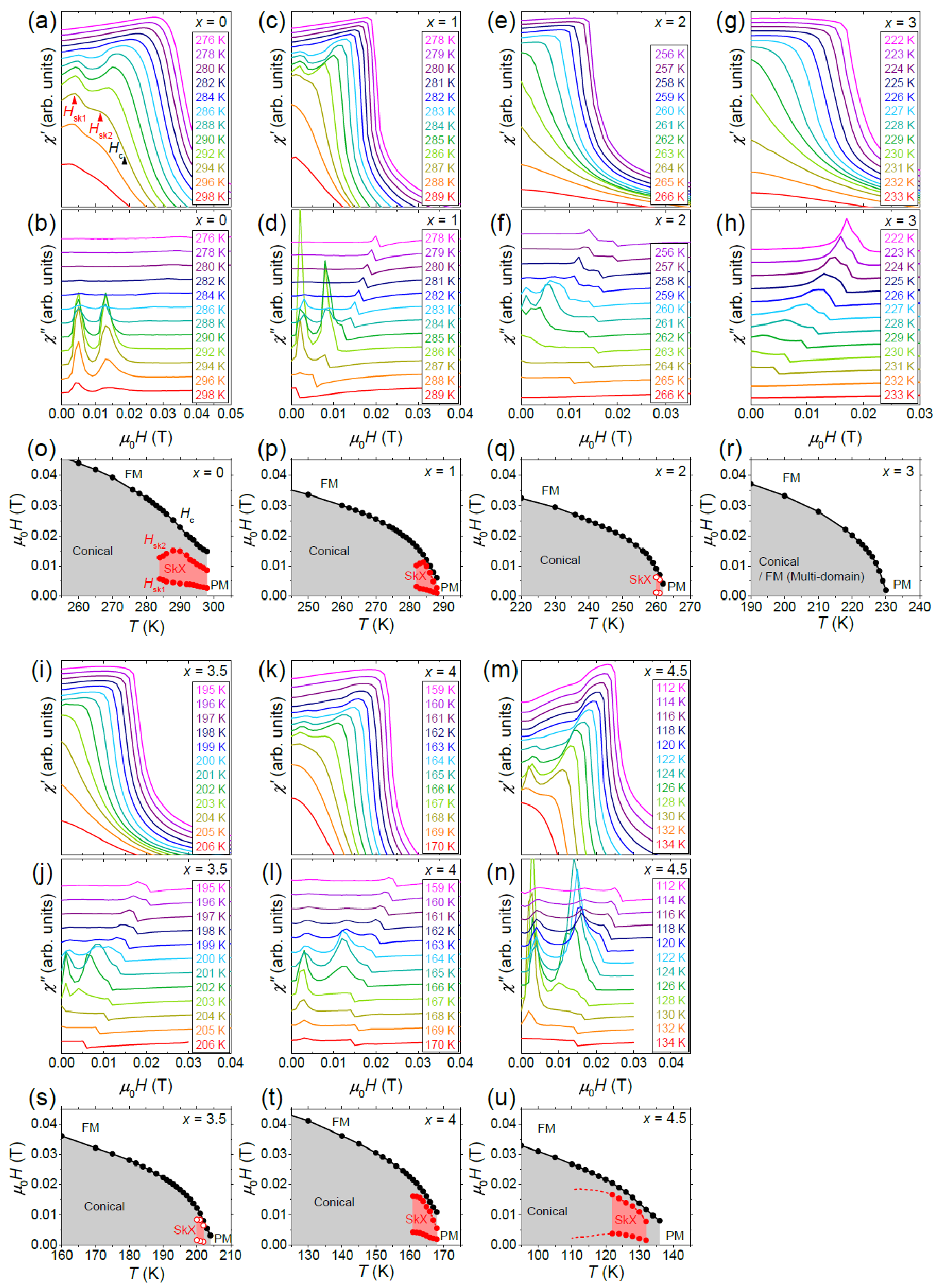}
\end{center}
\end{figure}

\newpage

\noindent
FIG. S4. (a-n) Magnetic field ($H$) dependence of the real part $\chi^\prime$ (upper panel) and the imaginary part $\chi^{\prime\prime}$ (lower panel) of the ac magnetic susceptibility in Co$_{8-x}$Fe$_x$Zn$_8$Mn$_4$, measured in a field-decreasing process from an induced-ferromagnetic region at several temperatures near $T_\mathrm{c}$. 
(o-u) $H$-$T$ phase diagram near $T_\mathrm{c}$ in Co$_{8-x}$Fe$_x$Zn$_8$Mn$_4$. 
For $x$ = 0, 1, 4 and 4.5, boundaries ($H_\mathrm{sk1}$ and $H_\mathrm{sk2}$) of skyrmion crystal (SkX) phase denoted with red closed circles are determined as the peak fields locating at both sides of dip structures in $\chi^\prime$, as indicated by red triangles in (a). 
Dotted red lines in $x$ = 4.5 indicate the fields where anomalies in $\chi^\prime$ are barely discernible, instead of clear dip structures observed near $T_\mathrm{c}$.
For $x$ = 2 and 3.5, SkX phase boundaries denoted with red open circles are determined as the fields where $\chi^{\prime\prime}$ exhibits double peaks. 
For all $x$, phase boundaries between conical (or ferromagnetic multi-domain for $x$ = 3) and induced-ferromagnetic phases denoted with black circles are determined as the inflection point of $\chi^\prime$ as indicated by a black triangle in (a). 

\newpage
\begin{figure}[htbp]
\begin{center}
\includegraphics[width=16cm]{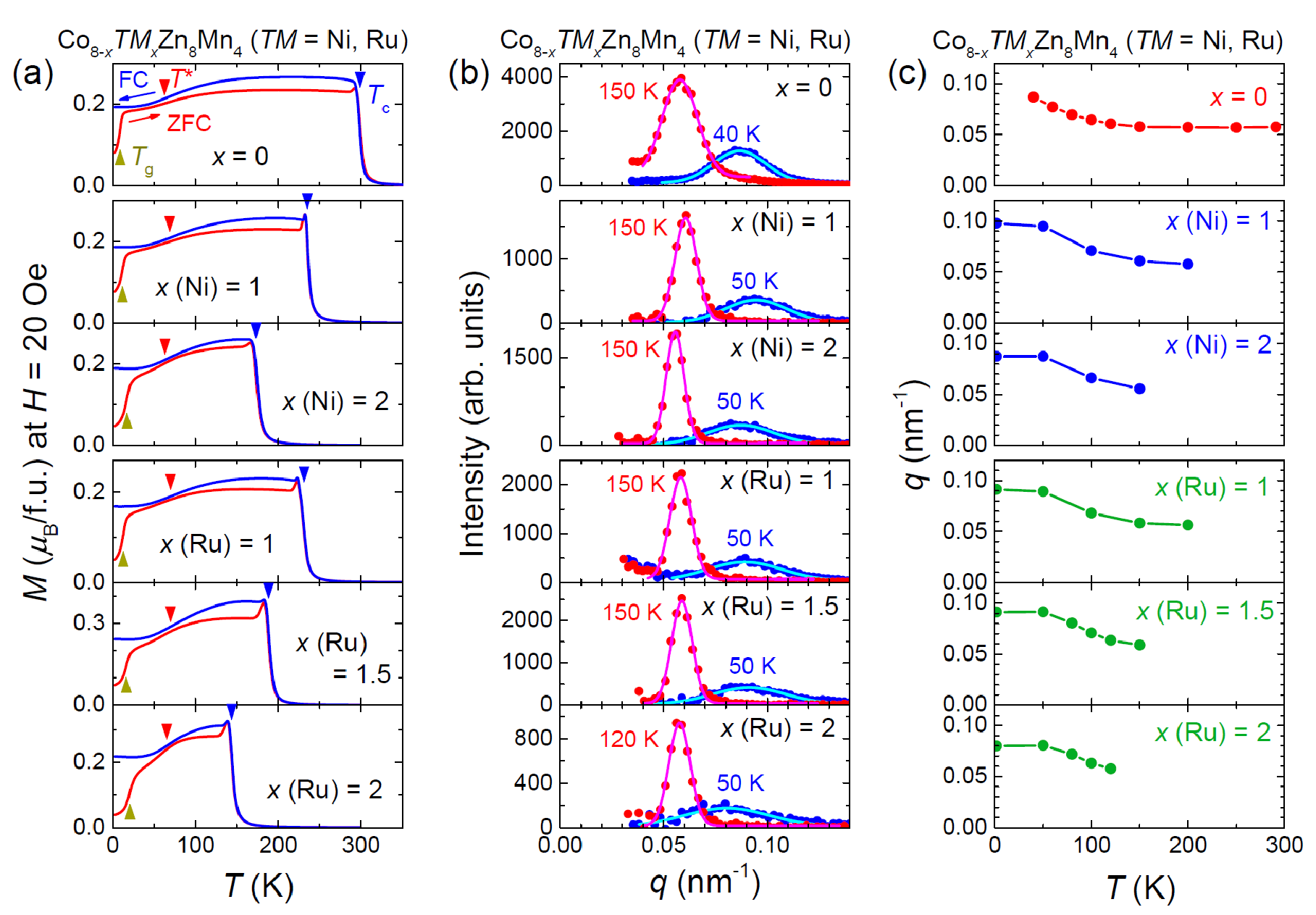}
\end{center}
\end{figure}
\noindent
FIG. S5. Magnetization and SANS data in Co$_{8-x}$Ni$_x$Zn$_8$Mn$_4$ (0 $\leq$ $x$ $\leq$ 2) and Co$_{8-x}$Ru$_x$Zn$_8$Mn$_4$ (0 $\leq$ $x$ $\leq$ 2). 
(a) Temperature dependence of magnetization under a magnetic field of 20 Oe. 
A blue line shows the data collected in a FC process, and a red line represents the data collected in a field-warming process after ZFC down to 2 K. 
$T_\mathrm{c}$ (blue triangle), $T_\mathrm{g}$ (yellow triangle) and $T^\ast$ (red triangles) are determined similarly to Fig. S3(a).
(b) Azimuthally-integrated SANS intensity under zero magnetic field is plotted as a function of $q$ at 150 K (120 K for $x$ = 2 in Ru doping) with red circles, and at 50 K (40 K for $x$ = 0) with blue circles. 
Data points are fitted to a Gaussian function (pink and light-blue lines). 
(c) Temperature dependence of helical $q$ value, determined as the peak center of the Gaussian function as plotted in (b).

\newpage

\begin{figure}[htbp]
\begin{center}
\includegraphics[width=9cm]{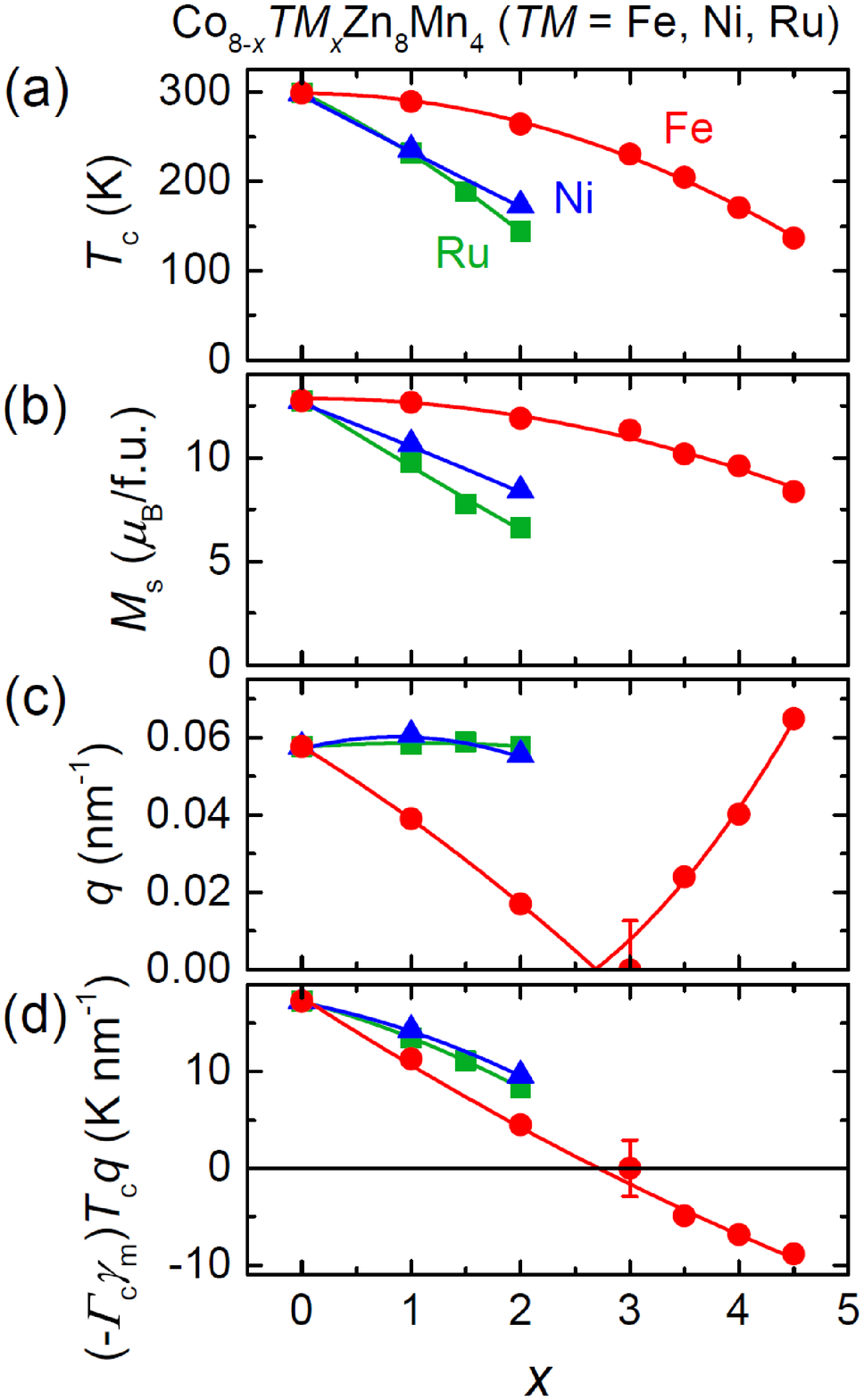}
\end{center}
\end{figure}
\noindent
FIG. S6. $x$ dependence of physical parameters in Co$_{8-x}$\textit{TM}$_x$Zn$_8$Mn$_4$ (\textit{TM} = Fe, Ni and Ru). 
(a) Helimagnetic transition temperature $T_\mathrm{c}$, determined by the temperature dependence of magnetization under 20 Oe.
(b) Saturation magnetization $M_\mathrm{s}$, obtained as a magnetization value at 2 K and 7 T.
(c) Helical $q$ value, determined by the Gaussian fitting to SANS intensity at 150 K (120 K for $x$ = 4.5 in Fe doping and $x$ = 2 in Ru doping) and at zero field.
(d) $\tilde{D} \equiv (-\Gamma_\mathrm{c}\gamma_\mathrm{m})T_\mathrm{c}q$, as a measure of Dzyaloshinskii-Moriya interaction. 
Sign of $\Gamma_\mathrm{c} \gamma_\mathrm{m}$ is determined by convergent-beam electron diffraction and Lorentz transmission electron microscopy measurements for $x$ = 0 and 4.5 (\textit{TM} = Fe), as shown in Fig. 4 in the main text. 

\newpage

\begin{figure}[htbp]
\begin{center}
\includegraphics[width=16cm]{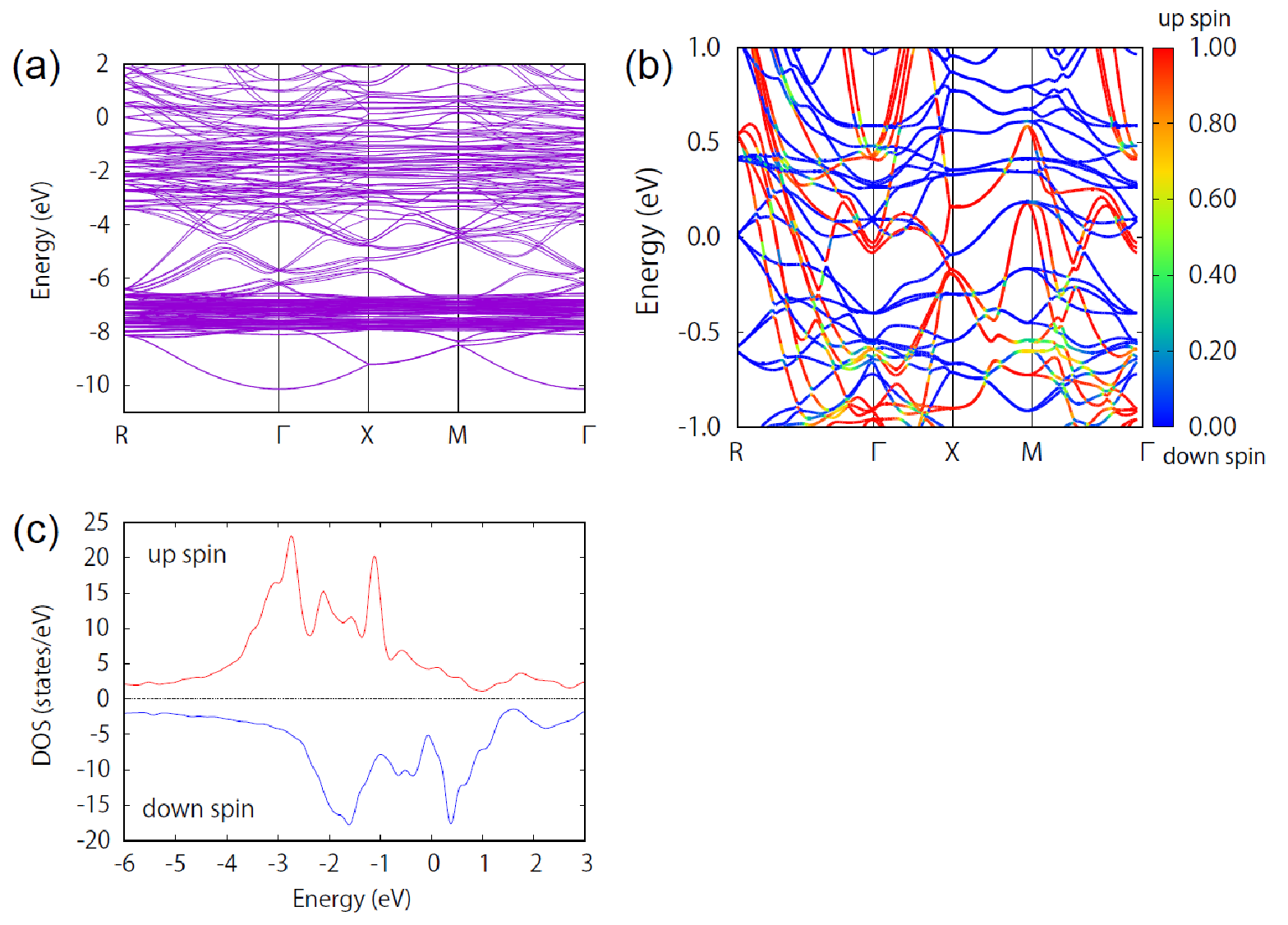}
\end{center}
\end{figure}
\noindent
FIG. S7. (a) Band structure of a hypothetical material Co$_{8}$Zn$_{12}$ in a ferromagnetic state, based on \textit{ab initio} density-functional theory that takes spin-orbit interaction into account. The Fermi level is set to the origin (0 eV).
(b) Expanded view of the band structure around the Fermi level. Color represents the weight of spin component, and its scale is displayed at right-side of the figure. 
(c) Density of states (DOS) for up-spin components (red line) and down-spin component (blue line).